\documentclass[pra,superscriptaddress,twocolumn,10pt,reprint]{revtex4-1}

\usepackage[utf8]{inputenc}
\usepackage{datetime}

\usepackage{graphicx}
\usepackage{tikz}
\usepackage{amsmath}
\usepackage{amssymb}
\usepackage{braket}
\usepackage[hidelinks]{hyperref}
\usepackage{printlen}
\usepackage{epstopdf}

\usepackage{hhline}

\newcommand{\del}[1]{}

\usepackage{newtxtext}
\usepackage{newtxmath}
\usepackage{microtype}

\graphicspath{{img/}}

\begin{document}
\title{From coherent collective excitation to Rydberg blockade on an atom chip}

\author{Julius de Hond}
\altaffiliation[Present address: ]{Research Laboratory of Electronics, MIT-Harvard Center for Ultracold Atoms, and Department of Physics, Massachusetts Institute of Technology, Cambridge, Massachusetts 02139, USA}
\affiliation{Van der Waals--Zeeman Institute, University of Amsterdam, Science Park 904, 1098 XH Amsterdam, The Netherlands}

\author{Rick van Bijnen}
\affiliation{Institut f\"ur Quantenoptik und Quanteninformation, \"Osterreichische Akademie der Wissenschaften, 6020 Innsbruck, Austria}

\author{S.J.J.M.F.\ Kokkelmans}
\affiliation{Eindhoven University of Technology, P.O. Box 513, 5600 MB Eindhoven, The Netherlands}

\author{R.J.C.\ Spreeuw}
\affiliation{Van der Waals--Zeeman Institute, University of Amsterdam, Science Park 904, 1098 XH Amsterdam, The Netherlands}

\author{H.B.\ van Linden van den Heuvell}
\affiliation{Van der Waals--Zeeman Institute, University of Amsterdam, Science Park 904, 1098 XH Amsterdam, The Netherlands}

\author{N.J.\ van Druten}
\email{n.j.vandruten@uva.nl}
\affiliation{Van der Waals--Zeeman Institute, University of Amsterdam, Science Park 904, 1098 XH Amsterdam, The Netherlands}

\date{this version compiled \today ,~\currenttime}

\begin{abstract}
Using time-resolved measurements, we demonstrate coherent collective Rydberg excitation crossing over into Rydberg blockade in a dense and ultracold gas trapped at a distance of 100~$\mu$m from a room-temperature atom chip. We perform Ramsey-type measurements to characterize the coherence. The experimental data are in good agreement with numerical results from a master equation using a mean-field approximation, and with results from a super-atom-based Hamiltonian. This represents significant progress in exploring a strongly interacting driven Rydberg gas on an atom chip.
\end{abstract}

\maketitle

\section{Introduction}
Rydberg atoms are being actively studied for their strong, tunable interactions, making them interesting tools for quantum information science, e.g.\ for quantum computing \cite{Saffman10,Saffman16} and quantum simulation \cite{Nguyen18,Schauss18} and for realizing strong photon-photon interactions \cite{Peyronel12,Hofmann13,Maxwell13,Baur14,Murray17,Thompson17,Busche17,Tiarks18}. Remarkable progress has been made both with ensembles \cite{Lukin01,Baur14,Ebert15,Balewski13} and with individual atoms \cite{Barredo14,Bernien17,Zeiher16}.
In dense samples the strong van-der-Waals interaction between Rydberg atoms limits excitation from the ground state to a single Rydberg excitation within some `blockade' volume. At the same time, atoms within this volume experience a coherent, collective coupling to the excitation light (as a `superatom')  with a Rabi frequency that is enhanced by $\sqrt{N_b}$, where $N_b$ is the number of participating atoms \cite{Low12}.
The coherence of the excitation competes with decoherence mechanisms such as spontaneous and black-body-induced decay of the Rydberg states \cite{Goldschmidt16,Young17,Boulier17}, atomic motion, the finite bandwidth of the excitation lasers, and, particularly in a trapped gas, with  dephasing through the inhomogeneous density distribution \cite{Dudin12,Low12,Whitlock16}.
Yet, collective enhancements of Rydberg excitations were observed on both one- \cite{Viteau11} and two-dimensional lattices \cite{Zeiher15}, in optical tweezer arrays \cite{Barredo14,Bernien17,Barredo18} and between physically separated ensembles that are individually dense enough to act collectively \cite{Ebert15}. 

\begin{figure}[t]
	\centering
	\begin{tikzpicture}
		\node[anchor = south west] (image) at (0,0) {\includegraphics[width=\columnwidth]{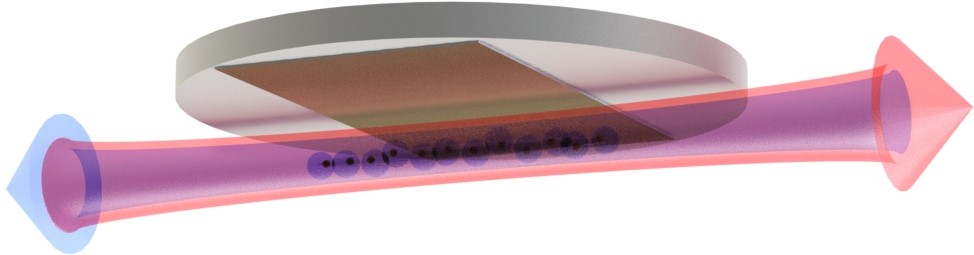}};
		\node[anchor = south west] at (0,2) {\fontfamily{cmr}\selectfont\large(a)};
	\end{tikzpicture}
	\hspace*{3pt}\includegraphics{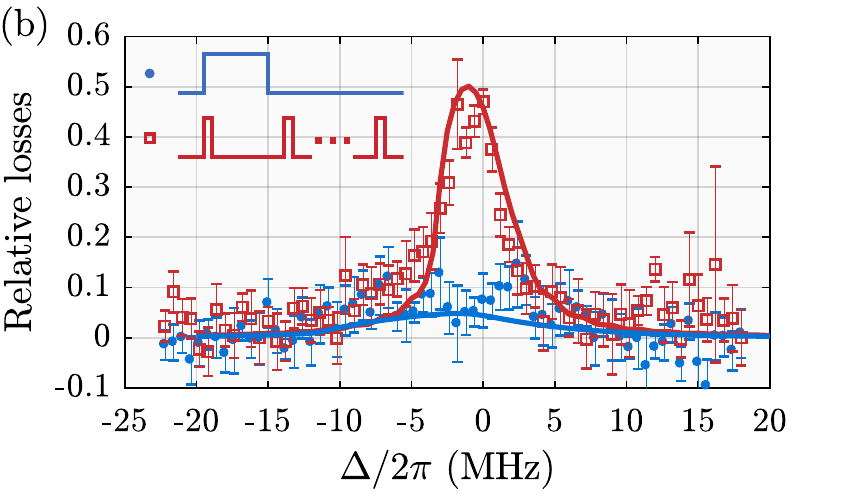}
	\caption{Demonstration of Rydberg blockade on an atom chip. (a) Illustration of our mounted chip trapping a cloud of atoms containing Rydberg excitations (black) in blockade spheres (blue). The two counter-propagating excitation beams are drawn as well. (b) Rydberg excitation of the $\ket{28D_{5/2}, m_J=5/2}$ state with two different pulse configurations with the same total pulse time, excited at a constant Rabi frequency of $\Omega/2\pi = 140~\mathrm{kHz}$. This dataset was obtained using $21\times 10^3$ atoms at a temperature of $T=0.7~\mathrm{\mu K}$. The open squares are a loss spectrum measured with 75 pulses of $500~\mathrm{ns}$ sent every $30~\mathrm{\mu s}$; the solid circles are a loss spectrum measured with a single $37.5~\mathrm{\mu s}$ pulse. Each point is the average of four measurements; the error bars denote one standard error of the mean. The solid lines are solutions to the mean-field master equation model (see the main text). Inset: schematic pulse shapes used for obtaining the spectra.}
	\label{fig:long-short-pulses}
\end{figure}

An ongoing challenge is to combine Rydberg-mediated interactions with atom chips \cite{Reichel11}, microfabricated devices designed to cool and trap neutral atoms close to surfaces. Such a combination would be an excellent starting point for hybrid quantum systems \cite{Petrosyan09,Xiang13}, e.g.\ combining solid-state and neutral-atom-based qubits, but also for building compact quantum devices based on neutral atoms.

However, thus far coherent Rydberg physics on atom chips has been difficult to observe due to decoherence from inhomogeneous Stark shifts induced by stray electric fields emanating from the surface \cite{Tauschinsky10,Hermann14,Cisternas17,Davtyan18}, which have also been observed and characterized close to a microwave waveguide \cite{Hogan12}, and using a chip in a cryogenic environment \cite{Thiele15}. While first evidence of Rydberg--Rydberg interactions on an atom chip has been demonstrated under cryogenic circumstances as well \cite{Teixeira15}, the saturation of Rydberg atom number could not be observed, and the excitation bandwidth was too large to probe the initial coherence.

Here, we demonstrate two key aspects of resonant Rydberg driving in a dense and ultracold gas, trapped at around 100~$\mu$m distance from a room-temperature atom chip. We describe how time-resolved measurements were used to detect (i) the early-time coherent regime of collectively enhanced excitations, crossing over at a later time into (ii) a regime with essentially constant Rydberg fraction due to the blockade mechanism. In regime (i) we find a collective enhancement of the Rydberg Rabi frequency by up to an order of magnitude compared to the single-atom case. This implies that there are up to 100 ground-state atoms per blockade volume in our system. In regime (ii) we have a Rydberg fraction much below unity (on the order of 1\%), consistent with the inferred blockade volume and despite the strong resonant driving of the system. In Sec.~\ref{sec:experimental} we describe these results, and in Sec.~\ref{sec:theory} we expand on the mean-field master equation and superatom models that were used to describe them, and find good agreement. In Sec.~\ref{sec:coherence}, finally, we describe a direct measurement of the excitation coherence using a Ramsey-type pulse sequence. These results constitute important progress in realizing coherent driving to a strongly interacting Rydberg gas  in an integrated quantum device, and in the presence of several decoherence mechanisms.

\section{Collective onset and Rydberg blockade}\label{sec:experimental}
The experimental arrangement is very similar to that described in Ref.~\cite{Cisternas17}. In short, we trap  $^{87}$Rb atoms in the $|F,m_F\rangle=|2,2\rangle$ state at a distance of $\sim\!100~\mathrm{\mu m}$ from the surface of an atom chip. The magnetic trap is generated with a $\mathsf{Z}$-shaped gold wire patterned on a silicon substrate \cite{vanEs10}. The temperature and atom number of the atomic cloud are controlled by varying the final frequency of the radio-frequency (RF) sweep used for evaporative cooling. This results in a cloud of typically $2\times 10^4$ atoms at a temperature of a few $\mathrm{\mu K}$ in the final magnetic trap with radial and axial trap frequencies of $\omega_\perp/2\pi = 860~\mathrm{Hz}$ and $\omega_\parallel/2\pi = 46~\mathrm{Hz}$, respectively, and with its trap bottom at $B_0 = 3.4~\mathrm{G}$. Thus, the atoms are confined in a highly elongated, cigar-shaped geometry with a typical width $\sigma_\perp = 2.6~\mathrm{\mu m}$ and length $\sigma_\parallel = 48~\mathrm{\mu m}$ (at $T = 2~\mathrm{\mu K}$); see Fig.~\ref{fig:long-short-pulses}(a).

For Rydberg excitation, we use a two-photon excitation scheme with a $780$-$\mathrm{nm}$ (`red') and a $480$-$\mathrm{nm}$-wavelength (`blue') laser. These propagate in opposite directions along the length of the cloud and have beam waists ($520~\mathrm{\mu m}$ and $90~\mathrm{\mu m}$, respectively) that are much larger than the width of the cloud, so that the (single-atom) Rabi frequency can be considered constant. We excite either the $\ket{28D_{5/2}, m_J = 5/2}$ or the $\ket{30S_{1/2}, m_J=1/2}$ state, see Table~\ref{tab:rydberg-parameters} for their relevant parameters. The red laser is blue detuned by $\Delta_r/2\pi = 100~\mathrm{MHz}$ from the intermediate $5P_{3/2}$ state to prevent off-resonant scattering from it. The blue laser is detuned by $\Delta_b \approx -\Delta_r$ such that the overall detuning $\Delta \equiv \Delta_r + \Delta_b$ is close to resonance. To change the overall detuning, we vary $\Delta_b$. The intermediate state can be adiabatically eliminated because $\Delta_r \gg \Omega_r,\Omega_b$ \cite{Brion07}, so that a two-level system with a Rabi frequency of $\Omega = \Omega_r\Omega_b/2\Delta_r$ remains (here $\Omega_r$ and $\Omega_b$ denote the red and blue Rabi frequency, respectively). 
The excitation lasers are stabilized using a reference cavity and have a linewidth of typically $10~\mathrm{kHz}$ \cite{deHond17}. The $780$-$\mathrm{nm}$ beam is sent through a combination of a mechanical shutter and an acousto-optical modulator for excellent contrast and fast switching times, respectively. 

Losses from the trap are detected via absorption imaging of the remaining atoms after a $6$-$\mathrm{ms}$ time of flight. A decrease in the atom number is interpreted as losses due to Rydberg excitation.
Most experiments described here are performed using a series of pulses (see below), with a pulse separation longer than the Rydberg lifetime. This allows us to investigate shorter timescales, where the losses per pulse are small, while still having an observable loss signal.

Due to their large orbitals, Rydberg atoms have a strong van-der-Waals interaction given by $-C_6/r^6$, where $r$ is the interatomic distance, and $C_6$ is the (state-dependent) effective interaction coefficient \cite{Gallagher94,Low12,Sibalic17,Weber17,Cisternas17}.
For typical nearest-neighbor distances in an ultracold ground-state gas the strength of this interaction is larger than the excitation bandwidth, $\zeta$, so it is not possible to excite a second Rydberg atom within some distance from a prior Rydberg excitation, leading to blockade. The relevant length scale over which Rydberg blockade is effective, known as the blockade radius \cite{Balewski14-2,Low12}, is in this case given by $\left| C_6/\hbar \zeta \right|^{1/6}$.
For narrow-bandwidth lasers (and as long as other decoherence mechanisms can be neglected) $\zeta$ is given by the generalized Rabi frequency $\sqrt{N_b\Omega^2 + \Delta^2}$, where the Rabi frequency is enhanced due to the collective coupling. For our parameters the blockade radius is typically more than $1~\mathrm{\mu m}$. Since the radial extent of our cloud is usually on the order of the blockade radius, the system is close to being 1D in terms of the geometrical extent of the Rydberg blockade, and we can only excite one Rydberg atom in the radial direction.

\begin{table}
	\begin{ruledtabular}
		\begin{tabular}{ccc}
				Rydberg state				&	$\ket{28D_{5/2}, m_J = 5/2}$	&	$\ket{30S_{1/2}, m_J = 1/2}$\\
				Lifetime ($\mathrm{\mu s}$)	&	$15.2$							&	$15.5$							\\
				$C_6/h$ ($\mathrm{MHz\,\mu m^6}$)&	$21.5$						&	$-25.0$
		\end{tabular}
	\end{ruledtabular}
	\caption{Overview of Rydberg state parameters as used in simulations. The $C_6$ coefficient of the $28D_{5/2}$ is anisotropic, we have tabulated its root-mean-square value after calculating it for different angles.}
	\label{tab:rydberg-parameters}
\end{table}

To demonstrate the presence of Rydberg blockade, we perform an experiment using the $\ket{28D_{5/2},m_J=5/2}$ state, see Fig.~\ref{fig:long-short-pulses}(b).
We compare the losses after a single $37.5$-$\mathrm{\mu s}$ pulse to those after a series of 75 pulses with a duration of $500~\mathrm{ns}$ apiece, in which case the pulse separation is longer than the Rydberg lifetime.
Ignoring the blockade and coherent behavior, one would naively expect these to lead to the same results since the pulse areas are the same. In Fig.~\ref{fig:long-short-pulses}(b) we can see that this is clearly not the case, the continuous pulse has far fewer losses compared to the pulse train. This can be interpreted in terms of the blockade strongly limiting the total Rydberg atom number during the continuous pulse. This effect is lifted when the pulse is split up in smaller segments, enhancing the losses.
The results are in very good agreement with a mean-field master equation model (see below). 

\begin{figure}
	\centering
	\includegraphics{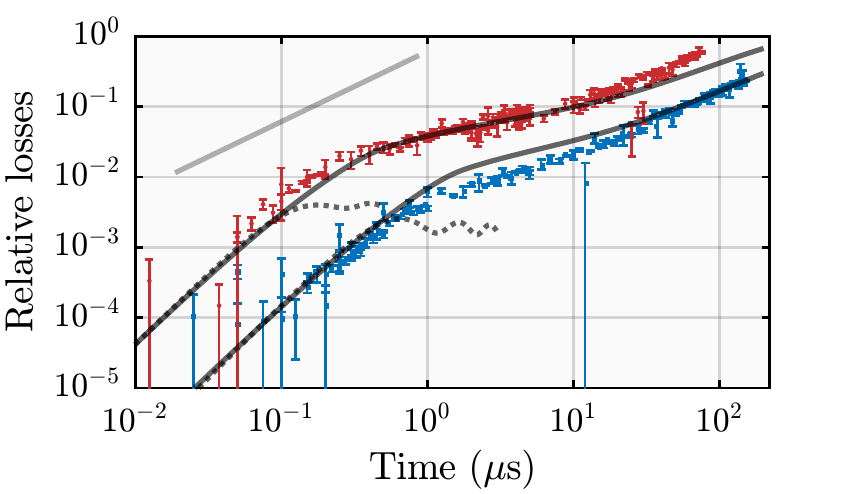}
	\caption{Regimes of collective Rydberg excitation observed in atom loss. Time evolution of the relative losses per pulse when exciting the $\ket{28D_{5/2}, m_J = 5/2}$ (red  circles, top data set) and $\ket{30S_{1/2}, m_J = 1/2}$ (blue squares, bottom data set) states resonantly. Using $N=2\times 10^4$ atoms at $T=2~\mathrm{\mu K}$ with two-photon, single-atom Rabi frequencies of $\Omega/2\pi \approx 210~\mathrm{kHz}$ and $40~\mathrm{kHz}$ for the $28D_{5/2}$ and $30S_{1/2}$ states, respectively. The error bars denote one standard error of the mean of three measurements. The solid and dotted lines are the result of the mean-field master equation model and the superatom Hamiltonian evolution, respectively. The solid, gray line indicates a linear relation, and serves to show deviations from this in both the coherent and in the blockaded regime.}
	\label{fig:30S-excitation-dynamics}
\end{figure}

More information can be extracted from the full time evolution while on resonance. For comparison, we look at the $\ket{30S_{1/2},m_J=1/2}$ state as well. The parameters of this state are similar to those of the $28D_{5/2}$ state, except its van-der-Waals interaction is isotropic and repulsive, also see Table~\ref{tab:rydberg-parameters} \cite{Sibalic17,Weber17}. For short times, we used multiple pulses and extracted the average losses per pulse afterwards, see Fig.~\ref{fig:30S-excitation-dynamics}.

For the single-atom Rabi frequencies used here (typically around $2\pi \times 100~\mathrm{kHz}$) three different regimes in the excitation evolution are clearly visible in Fig.~\ref{fig:30S-excitation-dynamics}. Specifically, we can distinguish between a coherent onset (where the scaling of losses with time is superlinear), a blockaded intermediate regime (where this scaling is sublinear), and a linear regime at later times. The latter is best understood as the regime where the cumulative number of atoms that has left the trap after decaying from the Rydberg state is larger than the number of atoms in the Rydberg state at any time. This lost atom number grows linearly since the process is governed by a (constant) decay rate. Note that for short pulse times, we are sensitive to losses below the $10^{-3}$ level (using ~500 pulses), i.e.\ to the loss of only a few atoms per pulse.

\section{Master equation and superatom model}\label{sec:theory}
The experimental results are explained well using a master-equation model following Ref.~\cite{Whitlock16}. The ground state and Rydberg state of each atom can be written as the spin-down and spin-up state of a pseudospin 1/2. The Rydberg--Rydberg interactions and the coherent coupling by the excitation lasers are then described by the following spin Hamiltonian:
\begin{equation}\label{eq:full-hamiltonian}
	\mathcal{H} = \sum\limits_{j = 1}^N \left( \frac{\hbar\Omega}{2} \sigma_x^{(j)} -\frac{\hbar\Delta}{2}\sigma_z^{(j)} + \mathcal{H}_\mathrm{int}^{(j)} \right),
\end{equation}
here $\sigma_{x/z}^{(i)}$ are the Pauli spin matrices of the $i^\mathrm{th}$ atom. The last term describes the Rydberg--Rydberg interactions, it is given by
\begin{equation}\label{eq:H-int-full}
	\mathcal{H}_\mathrm{int}^{(j)} = -\frac{C_6}{2} \sum_{k\neq j} \frac{P^{(j)} P^{(k)}}{ \left| x_j - x_k \right|^6}.
\end{equation}
Here $P^{(j)} \equiv \ket{r_j}\bra{r_j}$ is a projection operator onto the Rydberg state, and $x_{j/k}$ are the atom positions; the sum runs over all other atoms $k$. We assume the system to be in the frozen gas limit, meaning the displacement during the excitation is negligible. For excitation times longer than $\sim\!40~\mathrm{\mu s}$ this assumption is no longer tenable. However, we do not find this to be limiting the coherence. This is corroborated by our model using a single, constant decoherence rate (see below).

We can remove the complicated spatial correlations from the problem by replacing the interaction term in the Hamiltonian~(\ref{eq:full-hamiltonian}) with an interaction with the `mean field' of all other Rydberg excitations in the system (under the assumptions that there is at most one excitation per blockade volume). As shown in Ref.~\cite{Whitlock16}, this reduces Eq.~(\ref{eq:H-int-full}) to a single-atom term of the form
\begin{equation}
	\mathcal{H}_\mathrm{int}^{(j)} = V_j \sigma_z^{(j)}/2,
\end{equation}
where the mean-field interaction energy $V_j$ is given by:
\begin{equation}
	V_j = -C_6 \left(\frac{4\pi}{3} m n_0 \right)^{2}.
\end{equation}
This leads to the single-particle, mean-field Hamiltonian:
\begin{equation}
	\mathcal{H}_\mathrm{MF}^{(j)} = \frac{\hbar\Omega}{2}\sigma_x^{(j)}  -\frac{1}{2}\left( \hbar\Delta - V_j \right)\sigma_z^{(j)}.
\end{equation}
Here $m$ is the magnetization, which is nothing but the average Rydberg fraction. When we describe the system in terms of three levels (a ground state $\ket{g}$, a Rydberg state $\ket{r}$, and a loss state $\ket{l}$) this is equivalent to the $\rho_{rr}$ entry in the single-atom density matrix.

The single-atom, mean-field, Liouville--von Neumann master equation is now given by
\begin{equation}\label{eq:Lindblad-eq}
	\frac{\partial \rho^{(j)}}{\partial t} = -\frac{i}{\hbar} \left[ \mathcal{H}_\mathrm{MF}^{(j)}, \rho^{(j)} \right] + L \rho^{(j)} L^\dagger - \frac{1}{2} \left\{ L^\dagger L, \rho^{(j)} \right\} + \mathcal{L}_\mathrm{ph}^{(j)},
\end{equation}
here $\left\{\hat{A}, \hat{B}\right\} \equiv \hat{A}\hat{B} + \hat{B}\hat{A}$ denotes the anti-commutator, and $L$ is the jump operator describing spontaneous decay at a rate $\gamma$ from $\ket{r}$ into $\ket{g}$ and $\ket{l}$ with a branching ratio $b$:
\begin{equation}
	L = 
	\begin{pmatrix}
		0	&	\sqrt{\left( 1 - b\right) \gamma}	&	0\\
		0	&	0									&	0\\
		0	&	\sqrt{b\gamma}						&	0
	\end{pmatrix}.
\end{equation}
Our linewidth, $\gamma_0$ is generally limited by experimental factors such as stray fields or the finite laser linewidth. We add this using the phenomenological Lindblad operator
\begin{equation}
	\mathcal{L}_\mathrm{ph}^{(j)} = 
	\begin{pmatrix}
		0				&	-\gamma_0 \rho_{gr}/2		&	0	\\
		-\gamma_0 \rho_{rg}/2		&	0				&	0	\\
		0				&	0				&	0
	\end{pmatrix}.
\end{equation}

We numerically integrate Eq.~(\ref{eq:Lindblad-eq}) at different densities, and average the results over the cloud. The results [solid lines in Figs.~\ref{fig:long-short-pulses}(b) and \ref{fig:30S-excitation-dynamics}] are in good agreement with the experimental data once we include a decoherence rate on the order of $10^6~\mathrm{s^{-1}}$, see Table~\ref{tab:mastereq-parameters} for a complete account of the model parameters that were used.
In particular, the initial fast, superlinear rise in losses originating from the initial coherent Rydberg population buildup is reproduced, as well as the subsequent saturation of the Rydberg population, leading to a plateau in the relative losses as a function of time, until the cumulative losses via the Rydberg state start to dominate and the loss becomes linear with time. The key to observing the initial coherent rise is that while the \emph{single-atom} Rabi frequency is lower than the decoherence rate, the \emph{collective} Rabi frequency is enhanced (in our case by up to an order of magnitude) to the point that it dominates the short-time behavior.

\begin{table}
	\begin{ruledtabular}
		\begin{tabular}{lccc}
				&	State	&		$\Omega/2\pi$~(kHz)	&		$\gamma_0/2\pi~(10^6$ s$^{-1})$		\\\hline
			Fig.~\ref{fig:long-short-pulses}(b)		&	$28D_{5/2}$	&	140 	&	1 \\
			Fig.~\ref{fig:30S-excitation-dynamics}	&	$28D_{5/2}$	&	210 	&	4 \\
													&	$30S_{1/2}$	&	40 	&	1 \\
			Fig.~\ref{fig:ramsey-fringes-combined}(a)&	$30S_{1/2}$	&	50 	&	0.6\\
			Fig.~\ref{fig:ramsey-fringes-combined}(b)&	$30S_{1/2}$	&	105 	&	0.6
		\end{tabular}
	\end{ruledtabular}
	\caption{Overview of model parameters used in solving the mean-field master equation for all measurements presented in the main text.}
	\label{tab:mastereq-parameters}
\end{table}

We have also simulated the system using the full many-body Hamiltonian describing the excitation of interacting Rydberg atoms, whilst ignoring inelastic and decohering processes. 
The Rydberg blockade can be used to curb the otherwise unwieldy size of the basis set: it prohibits a large fraction of basis states to be populated, meaning the calculation can be simplified dramatically. We do so by dividing the cloud in so called `superatoms': entities consisting of multiple ground-state atoms that couple to the light field collectively \cite{vanBijnen11,Robicheaux05}. Furthermore, we restrict ourselves to a one-dimensional description.
This way, it is computationally feasible to propagate the Hamiltonian over the first few $\mathrm{\mu s}$ which is sufficient to analyze the coherent start.
The results are shown as dotted lines in Fig.~\ref{fig:30S-excitation-dynamics}. These confirm the short-time results of the mean-field model. For intermediate times (after the initial fast rise), the dotted lines show how the inhomogeneous density along the length of the cloud leads to a dephasing of the collective Rabi oscillations at various positions in the cloud, even in the absence of other decoherence mechanisms. The difference in saturation level between the mean-field and fully coherent models can be (partially) attributed to the fact that the latter is strictly one-dimensional, while the mean-field model does not impose any restrictions on the number of Rydberg excitations that can be present in the radial direction.

The differences that we observe between the $S$ and $D$ state can be explained in terms of the different Rabi frequencies used during the excitation. The fact that the $28D_{5/2}$ state has an anisotropic interaction does not appear to matter much, as our models use a root-mean-square averaged value and still produce a good result. This can be understood by realizing our cloud has a finite radial extent, meaning the interaction will be averaged over the different positions atoms can take with respect to one another.

\section{Direct measurement of coherence}\label{sec:coherence}
\begin{figure}
	\centering
	\includegraphics{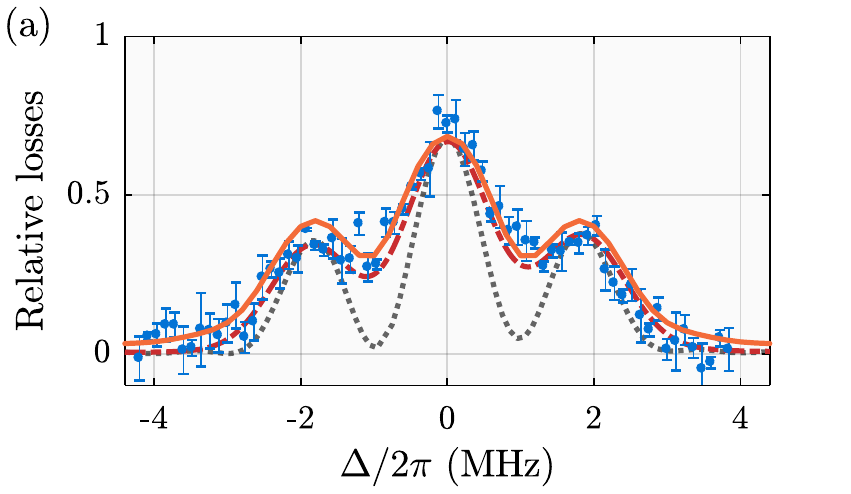}\\%
	\includegraphics{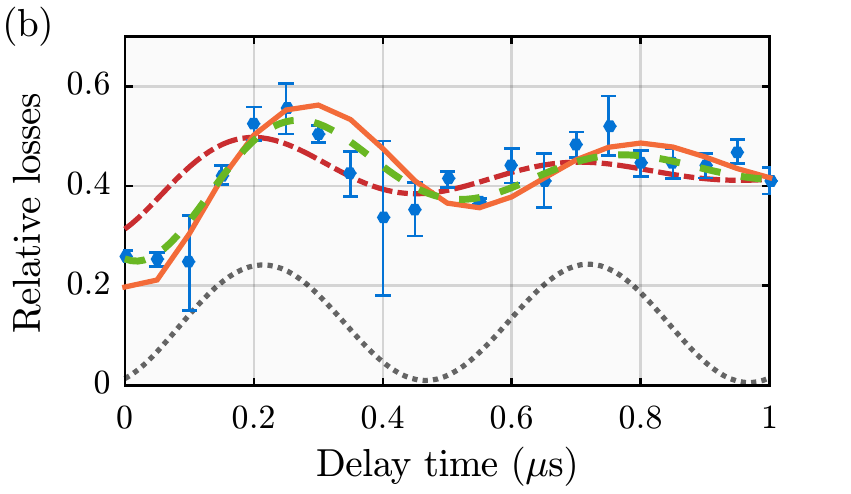}
	\caption{Ramsey fringes measured both in (a) the frequency and (b) time domain to assess the coherence of the excitation of the $\ket{30S_{1/2}, m_J = 1/2}$ state. In both measurements (blue circles) we used multiple pair pulses sent at intervals longer than the Rydberg lifetime. The pairs consisted of two $250$-$\mathrm{ns}$ pulses, either at a variable detuning (a, time between pulses fixed to $250~\mathrm{ns}$, measured using $13\times 10^3$ atoms at $T = 2~\mathrm{\mu K}$) or using a variable time between the pulses (b, detuning fixed to $\Delta/2\pi = -2~\mathrm{MHz}$, measured using $N = 10^4$ atoms at $T = 1~\mathrm{\mu K}$). The orange, solid lines are the results of the mean-field master equation. The gray, dotted lines are the populations in the Rydberg state obtained by evolving the Hamiltonian as described in the main text; this is either convolved with a Gaussian (a) or given an offset and an exponentially decaying envelope (b) to approximate the effects of decoherence (dash-dotted, red lines). The bottom figure also contains a fit to an exponentially decaying sine (dashed, green line), from which we extract a coherence time of $1/\gamma = 0.44~\mathrm{\mu s}$.}
	\label{fig:ramsey-fringes-combined}
\end{figure}

In order to demonstrate the coherence explicitly, we have performed Ramsey-type experiments. We employed a double-pulse scheme, where we varied the detuning and used two pulses of $250~\mathrm{ns}$ with a $250~\mathrm{ns}$ gap in between them. To increase the signal we repeated such a pulse pair 300 times. The observed Ramsey fringes are shown in Fig.~\ref{fig:ramsey-fringes-combined}(a). The results of the mean-field approach (solid lines in Fig.~\ref{fig:ramsey-fringes-combined}) are in excellent agreement with our measurements. On the other hand, the coherent propagation of the Hamiltonian shows the same qualitative behavior, with much stronger contrast. This can be attributed to the absence of decoherence in that model. 

Convolving the calculated spectrum with a Gaussian effectively simulates inhomogeneous spectral broadening. In this way, we
find an effective linewidth $\sigma=0.42~\mathrm{MHz}$.
Finally, we measure the decoherence time directly, using a Ramsey measurement in the time domain. To this end, we detune the lasers to the red by $\Delta/2\pi = -2~\mathrm{MHz}$, and again use two $250~\mathrm{ns}$ pulses, but now vary the delay time between them. Repeating the pulse pair 100 times, we observe a decaying sine, see Fig.~\ref{fig:ramsey-fringes-combined}(b). From a fit to this we can extract a coherence time of $1/\gamma = 0.44~\mathrm{\mu s}$. This is consistent with the linewidth $\sigma$ mentioned above (since $\sigma/\gamma \approx 1/2\pi$).

Due to dephasing, there are some losses for zero delay times which are not reproduced by the coherent Hamiltonian model. Here, we have mimicked decoherence by adding an exponentially decaying envelope and an offset to account for the nonzero losses at zero delay times. Note that the amplitude of the oscillation is not a fit parameter and matches the experimental results

We now turn to the source of the observed decoherence and dephasing. The dominant cause is likely the presence of inhomogeneous electric fields, leading to observed linewidths on the order of $1~\mathrm{MHz}$ in our system, via the inhomogeneous Stark shift of the Rydberg states involved \cite{Cisternas17}.
The static, homogeneous part of this field has been stable over the course of a few years, but there is of course also the possibility of electric field noise close to the surface of our chip; at present we cannot discriminate between this noise and a constant gradient in the electric field over the length of our cloud as the source of dephasing.
Apart from that, there is dephasing via coupling to black-body-populated nearby Rydberg states \cite{Boulier17,Young17,Goldschmidt16}, and there is some residual motion due to the finite temperature. Combined, these explain why our measurements do not reflect the weak Rabi oscillations predicted by the fully coherent evolution (see e.g.\ Fig.~\ref{fig:30S-excitation-dynamics}), in which the dephasing is dominated by the inhomogeneous density. On longer timescales than those probed here, the black-body-induced resonant dipole--dipole interaction does lead to stronger dephasing; we have probed this regime as well and find that our measurements can be explained with an extended version of the mean-field model \cite{forthcomingpaper}.

\section{Conclusions}
In conclusion, we have directly observed blockade and coherence in the excitation of Rydberg atoms in a mesoscopic ensemble of atoms trapped on an atom chip. We have analyzed the measured spectra and the dynamical evolution both by solving the many-body Hamiltonian and a mean-field master equation. We are able to characterize and quantify the amount of decoherence, which we attribute to inhomogeneous fields resulting from the present chip design.

In the future, finer control of the electric field components can be gained by adding more on-chip electrodes. This would allow compensating electric field gradients
and significantly extend the coherence time. In parallel, adding an optical lattice along the length of the cloud (as in Ref.~\cite{Dudin12}) will turn the system into a linear array of tens to hundreds of atomic ensembles on a chip, with the distance between sites smaller than the characteristic blockade radius. An intriguing further possibility would then be to use an electric field gradient along the length of the cloud to allow for site-specific control over the excitation within such an array. Furthermore, it should be relatively straightforward to extend these results to colder and denser clouds and Bose-Einstein condensates, to reach more deeply into the one-dimensional regime where the radial cloud size is significantly smaller than the blockade radius. It would also be interesting to extend the Ramsey sequence used here to more advanced (spin-echo-like) pulse sequences, in order to further distinguish and characterize the various dephasing and decoherence mechanisms mentioned above. This would also allow us to discriminate between (constant) gradients in the electric field and contributions from electric field noise.

The results presented here open up promising routes towards quantum information science with Rydberg atoms in integrated quantum devices.

\section*{Acknowledgments}
We thank Shannon Whitlock and Tilman Pfau for fruitful discussions and suggestions. R.J.C.S.\ and N.J.v.D.\ acknowledge stimulating discussions within the QuSoft consortium. This work is part of the research programme of the Foundation for Fundamental Research on Matter (FOM), which is part of the Dutch Organisation for Scientific Research (NWO). We also acknowledge EU funding through the RySQ programme. The superatom calculations were carried out on the Lisa Compute Cluster, which is hosted at SURFSara (\url{www.surfsara.nl}).

\bibliography{reduced}

\begin{thebibliography}{47}%
\makeatletter
\providecommand \@ifxundefined [1]{%
 \@ifx{#1\undefined}
}%
\providecommand \@ifnum [1]{%
 \ifnum #1\expandafter \@firstoftwo
 \else \expandafter \@secondoftwo
 \fi
}%
\providecommand \@ifx [1]{%
 \ifx #1\expandafter \@firstoftwo
 \else \expandafter \@secondoftwo
 \fi
}%
\providecommand \natexlab [1]{#1}%
\providecommand \enquote  [1]{``#1''}%
\providecommand \bibnamefont  [1]{#1}%
\providecommand \bibfnamefont [1]{#1}%
\providecommand \citenamefont [1]{#1}%
\providecommand \href@noop [0]{\@secondoftwo}%
\providecommand \href [0]{\begingroup \@sanitize@url \@href}%
\providecommand \@href[1]{\@@startlink{#1}\@@href}%
\providecommand \@@href[1]{\endgroup#1\@@endlink}%
\providecommand \@sanitize@url [0]{\catcode `\\12\catcode `\$12\catcode
  `\&12\catcode `\#12\catcode `\^12\catcode `\_12\catcode `\%12\relax}%
\providecommand \@@startlink[1]{}%
\providecommand \@@endlink[0]{}%
\providecommand \url  [0]{\begingroup\@sanitize@url \@url }%
\providecommand \@url [1]{\endgroup\@href {#1}{\urlprefix }}%
\providecommand \urlprefix  [0]{URL }%
\providecommand \Eprint [0]{\href }%
\providecommand \doibase [0]{http://dx.doi.org/}%
\providecommand \selectlanguage [0]{\@gobble}%
\providecommand \bibinfo  [0]{\@secondoftwo}%
\providecommand \bibfield  [0]{\@secondoftwo}%
\providecommand \translation [1]{[#1]}%
\providecommand \BibitemOpen [0]{}%
\providecommand \bibitemStop [0]{}%
\providecommand \bibitemNoStop [0]{.\EOS\space}%
\providecommand \EOS [0]{\spacefactor3000\relax}%
\providecommand \BibitemShut  [1]{\csname bibitem#1\endcsname}%
\let\auto@bib@innerbib\@empty
\bibitem [{\citenamefont {Saffman}\ \emph {et~al.}(2010)\citenamefont
  {Saffman}, \citenamefont {Walker},\ and\ \citenamefont
  {M{\o}lmer}}]{Saffman10}%
  \BibitemOpen
  \bibfield  {author} {\bibinfo {author} {\bibfnamefont {M.}~\bibnamefont
  {Saffman}}, \bibinfo {author} {\bibfnamefont {T.~G.}\ \bibnamefont {Walker}},
  \ and\ \bibinfo {author} {\bibfnamefont {K.}~\bibnamefont {M{\o}lmer}},\
  }\href@noop {} {\bibfield  {journal} {\bibinfo  {journal} {Reviews of Modern
  Physics}\ }\textbf {\bibinfo {volume} {82}},\ \bibinfo {pages} {2313}
  (\bibinfo {year} {2010})}\BibitemShut {NoStop}%
\bibitem [{\citenamefont {Saffman}(2016)}]{Saffman16}%
  \BibitemOpen
  \bibfield  {author} {\bibinfo {author} {\bibfnamefont {M.}~\bibnamefont
  {Saffman}},\ }\href {http://stacks.iop.org/0953-4075/49/i=20/a=202001}
  {\bibfield  {journal} {\bibinfo  {journal} {Journal of Physics B: Atomic,
  Molecular and Optical Physics}\ }\textbf {\bibinfo {volume} {49}},\ \bibinfo
  {pages} {202001} (\bibinfo {year} {2016})}\BibitemShut {NoStop}%
\bibitem [{\citenamefont {Nguyen}\ \emph {et~al.}(2018)\citenamefont {Nguyen},
  \citenamefont {Raimond}, \citenamefont {Sayrin}, \citenamefont {Corti\~nas},
  \citenamefont {Cantat-Moltrecht}, \citenamefont {Assemat}, \citenamefont
  {Dotsenko}, \citenamefont {Gleyzes}, \citenamefont {Haroche}, \citenamefont
  {Roux}, \citenamefont {Jolicoeur},\ and\ \citenamefont {Brune}}]{Nguyen18}%
  \BibitemOpen
  \bibfield  {author} {\bibinfo {author} {\bibfnamefont {T.~L.}\ \bibnamefont
  {Nguyen}}, \bibinfo {author} {\bibfnamefont {J.~M.}\ \bibnamefont {Raimond}},
  \bibinfo {author} {\bibfnamefont {C.}~\bibnamefont {Sayrin}}, \bibinfo
  {author} {\bibfnamefont {R.}~\bibnamefont {Corti\~nas}}, \bibinfo {author}
  {\bibfnamefont {T.}~\bibnamefont {Cantat-Moltrecht}}, \bibinfo {author}
  {\bibfnamefont {F.}~\bibnamefont {Assemat}}, \bibinfo {author} {\bibfnamefont
  {I.}~\bibnamefont {Dotsenko}}, \bibinfo {author} {\bibfnamefont
  {S.}~\bibnamefont {Gleyzes}}, \bibinfo {author} {\bibfnamefont
  {S.}~\bibnamefont {Haroche}}, \bibinfo {author} {\bibfnamefont
  {G.}~\bibnamefont {Roux}}, \bibinfo {author} {\bibfnamefont {T.}~\bibnamefont
  {Jolicoeur}}, \ and\ \bibinfo {author} {\bibfnamefont {M.}~\bibnamefont
  {Brune}},\ }\href {\doibase 10.1103/PhysRevX.8.011032} {\bibfield  {journal}
  {\bibinfo  {journal} {Phys. Rev. X}\ }\textbf {\bibinfo {volume} {8}},\
  \bibinfo {pages} {011032} (\bibinfo {year} {2018})}\BibitemShut {NoStop}%
\bibitem [{\citenamefont {{Schauss}}(2018)}]{Schauss18}%
  \BibitemOpen
  \bibfield  {author} {\bibinfo {author} {\bibfnamefont {P.}~\bibnamefont
  {{Schauss}}},\ }\href {\doibase 10.1088/2058-9565/aa9c59} {\bibfield
  {journal} {\bibinfo  {journal} {Quantum Science and Technology}\ }\textbf
  {\bibinfo {volume} {3}},\ \bibinfo {pages} {023001} (\bibinfo {year}
  {2018})}\BibitemShut {NoStop}%
\bibitem [{\citenamefont {Peyronel}\ \emph {et~al.}(2012)\citenamefont
  {Peyronel}, \citenamefont {Firstenberg}, \citenamefont {Liang}, \citenamefont
  {Hofferbeth}, \citenamefont {Gorshkov}, \citenamefont {Pohl}, \citenamefont
  {Lukin},\ and\ \citenamefont {Vuleti\'c}}]{Peyronel12}%
  \BibitemOpen
  \bibfield  {author} {\bibinfo {author} {\bibfnamefont {T.}~\bibnamefont
  {Peyronel}}, \bibinfo {author} {\bibfnamefont {O.}~\bibnamefont
  {Firstenberg}}, \bibinfo {author} {\bibfnamefont {Q.-Y.}\ \bibnamefont
  {Liang}}, \bibinfo {author} {\bibfnamefont {S.}~\bibnamefont {Hofferbeth}},
  \bibinfo {author} {\bibfnamefont {A.~V.}\ \bibnamefont {Gorshkov}}, \bibinfo
  {author} {\bibfnamefont {T.}~\bibnamefont {Pohl}}, \bibinfo {author}
  {\bibfnamefont {M.~D.}\ \bibnamefont {Lukin}}, \ and\ \bibinfo {author}
  {\bibfnamefont {V.}~\bibnamefont {Vuleti\'c}},\ }\href@noop {} {\bibfield
  {journal} {\bibinfo  {journal} {Nature}\ }\textbf {\bibinfo {volume} {488}},\
  \bibinfo {pages} {57} (\bibinfo {year} {2012})}\BibitemShut {NoStop}%
\bibitem [{\citenamefont {Hofmann}\ \emph {et~al.}(2013)\citenamefont
  {Hofmann}, \citenamefont {G\"unter}, \citenamefont {Schempp}, \citenamefont
  {Robert-de Saint-Vincent}, \citenamefont {G\"arttner}, \citenamefont {Evers},
  \citenamefont {Whitlock},\ and\ \citenamefont {Weidem\"uller}}]{Hofmann13}%
  \BibitemOpen
  \bibfield  {author} {\bibinfo {author} {\bibfnamefont {C.~S.}\ \bibnamefont
  {Hofmann}}, \bibinfo {author} {\bibfnamefont {G.}~\bibnamefont {G\"unter}},
  \bibinfo {author} {\bibfnamefont {H.}~\bibnamefont {Schempp}}, \bibinfo
  {author} {\bibfnamefont {M.}~\bibnamefont {Robert-de Saint-Vincent}},
  \bibinfo {author} {\bibfnamefont {M.}~\bibnamefont {G\"arttner}}, \bibinfo
  {author} {\bibfnamefont {J.}~\bibnamefont {Evers}}, \bibinfo {author}
  {\bibfnamefont {S.}~\bibnamefont {Whitlock}}, \ and\ \bibinfo {author}
  {\bibfnamefont {M.}~\bibnamefont {Weidem\"uller}},\ }\href {\doibase
  10.1103/PhysRevLett.110.203601} {\bibfield  {journal} {\bibinfo  {journal}
  {Physical Review Letters}\ }\textbf {\bibinfo {volume} {110}},\ \bibinfo
  {pages} {203601} (\bibinfo {year} {2013})}\BibitemShut {NoStop}%
\bibitem [{\citenamefont {Maxwell}\ \emph {et~al.}(2013)\citenamefont
  {Maxwell}, \citenamefont {Szwer}, \citenamefont {Paredes-Barato},
  \citenamefont {Busche}, \citenamefont {Pritchard}, \citenamefont {Gauguet},
  \citenamefont {Weatherill}, \citenamefont {Jones},\ and\ \citenamefont
  {Adams}}]{Maxwell13}%
  \BibitemOpen
  \bibfield  {author} {\bibinfo {author} {\bibfnamefont {D.}~\bibnamefont
  {Maxwell}}, \bibinfo {author} {\bibfnamefont {D.~J.}\ \bibnamefont {Szwer}},
  \bibinfo {author} {\bibfnamefont {D.}~\bibnamefont {Paredes-Barato}},
  \bibinfo {author} {\bibfnamefont {H.}~\bibnamefont {Busche}}, \bibinfo
  {author} {\bibfnamefont {J.~D.}\ \bibnamefont {Pritchard}}, \bibinfo {author}
  {\bibfnamefont {A.}~\bibnamefont {Gauguet}}, \bibinfo {author} {\bibfnamefont
  {K.~J.}\ \bibnamefont {Weatherill}}, \bibinfo {author} {\bibfnamefont
  {M.~P.~A.}\ \bibnamefont {Jones}}, \ and\ \bibinfo {author} {\bibfnamefont
  {C.~S.}\ \bibnamefont {Adams}},\ }\href {\doibase
  10.1103/PhysRevLett.110.103001} {\bibfield  {journal} {\bibinfo  {journal}
  {Physical Review Letters}\ }\textbf {\bibinfo {volume} {110}},\ \bibinfo
  {pages} {103001} (\bibinfo {year} {2013})}\BibitemShut {NoStop}%
\bibitem [{\citenamefont {Baur}\ \emph {et~al.}(2014)\citenamefont {Baur},
  \citenamefont {Tiarks}, \citenamefont {Rempe},\ and\ \citenamefont
  {D\"urr}}]{Baur14}%
  \BibitemOpen
  \bibfield  {author} {\bibinfo {author} {\bibfnamefont {S.}~\bibnamefont
  {Baur}}, \bibinfo {author} {\bibfnamefont {D.}~\bibnamefont {Tiarks}},
  \bibinfo {author} {\bibfnamefont {G.}~\bibnamefont {Rempe}}, \ and\ \bibinfo
  {author} {\bibfnamefont {S.}~\bibnamefont {D\"urr}},\ }\href {\doibase
  10.1103/PhysRevLett.112.073901} {\bibfield  {journal} {\bibinfo  {journal}
  {Physical Review Letters}\ }\textbf {\bibinfo {volume} {112}},\ \bibinfo
  {pages} {073901} (\bibinfo {year} {2014})}\BibitemShut {NoStop}%
\bibitem [{\citenamefont {Murray}\ and\ \citenamefont {Pohl}(2017)}]{Murray17}%
  \BibitemOpen
  \bibfield  {author} {\bibinfo {author} {\bibfnamefont {C.~R.}\ \bibnamefont
  {Murray}}\ and\ \bibinfo {author} {\bibfnamefont {T.}~\bibnamefont {Pohl}},\
  }\href@noop {} {\bibfield  {journal} {\bibinfo  {journal} {Physical Review
  X}\ }\textbf {\bibinfo {volume} {7}},\ \bibinfo {pages} {031007} (\bibinfo
  {year} {2017})}\BibitemShut {NoStop}%
\bibitem [{\citenamefont {Thompson}\ \emph {et~al.}(2017)\citenamefont
  {Thompson}, \citenamefont {Nicholson}, \citenamefont {Liang}, \citenamefont
  {Cantu}, \citenamefont {Venkatramani}, \citenamefont {Choi}, \citenamefont
  {Fedorov}, \citenamefont {Viscor}, \citenamefont {Pohl}, \citenamefont
  {Lukin},\ and\ \citenamefont {Vuleti\'c}}]{Thompson17}%
  \BibitemOpen
  \bibfield  {author} {\bibinfo {author} {\bibfnamefont {J.~D.}\ \bibnamefont
  {Thompson}}, \bibinfo {author} {\bibfnamefont {T.}~\bibnamefont {Nicholson}},
  \bibinfo {author} {\bibfnamefont {Q.-Y.}\ \bibnamefont {Liang}}, \bibinfo
  {author} {\bibfnamefont {S.~H.}\ \bibnamefont {Cantu}}, \bibinfo {author}
  {\bibfnamefont {A.~V.}\ \bibnamefont {Venkatramani}}, \bibinfo {author}
  {\bibfnamefont {S.}~\bibnamefont {Choi}}, \bibinfo {author} {\bibfnamefont
  {I.~A.}\ \bibnamefont {Fedorov}}, \bibinfo {author} {\bibfnamefont
  {D.}~\bibnamefont {Viscor}}, \bibinfo {author} {\bibfnamefont
  {T.}~\bibnamefont {Pohl}}, \bibinfo {author} {\bibfnamefont {M.~D.}\
  \bibnamefont {Lukin}}, \ and\ \bibinfo {author} {\bibfnamefont
  {V.}~\bibnamefont {Vuleti\'c}},\ }\href@noop {} {\bibfield  {journal}
  {\bibinfo  {journal} {Nature}\ }\textbf {\bibinfo {volume} {542}},\ \bibinfo
  {pages} {206} (\bibinfo {year} {2017})}\BibitemShut {NoStop}%
\bibitem [{\citenamefont {Busche}\ \emph {et~al.}(2017)\citenamefont {Busche},
  \citenamefont {Huillery}, \citenamefont {Ball}, \citenamefont {Ilieva},
  \citenamefont {Jones},\ and\ \citenamefont {Adams}}]{Busche17}%
  \BibitemOpen
  \bibfield  {author} {\bibinfo {author} {\bibfnamefont {H.}~\bibnamefont
  {Busche}}, \bibinfo {author} {\bibfnamefont {P.}~\bibnamefont {Huillery}},
  \bibinfo {author} {\bibfnamefont {S.~W.}\ \bibnamefont {Ball}}, \bibinfo
  {author} {\bibfnamefont {T.}~\bibnamefont {Ilieva}}, \bibinfo {author}
  {\bibfnamefont {M.~P.~A.}\ \bibnamefont {Jones}}, \ and\ \bibinfo {author}
  {\bibfnamefont {C.~S.}\ \bibnamefont {Adams}},\ }\href@noop {} {\bibfield
  {journal} {\bibinfo  {journal} {Nature Physics}\ }\textbf {\bibinfo {volume}
  {13}},\ \bibinfo {pages} {655} (\bibinfo {year} {2017})}\BibitemShut
  {NoStop}%
\bibitem [{\citenamefont {Tiarks}\ \emph {et~al.}(2018)\citenamefont {Tiarks},
  \citenamefont {Schmidt-Eberle}, \citenamefont {Stolz}, \citenamefont
  {Rempe},\ and\ \citenamefont {D\"{u}rr}}]{Tiarks18}%
  \BibitemOpen
  \bibfield  {author} {\bibinfo {author} {\bibfnamefont {D.}~\bibnamefont
  {Tiarks}}, \bibinfo {author} {\bibfnamefont {S.}~\bibnamefont
  {Schmidt-Eberle}}, \bibinfo {author} {\bibfnamefont {T.}~\bibnamefont
  {Stolz}}, \bibinfo {author} {\bibfnamefont {G.}~\bibnamefont {Rempe}}, \ and\
  \bibinfo {author} {\bibfnamefont {S.}~\bibnamefont {D\"{u}rr}},\ }\href@noop
  {} {\bibfield  {journal} {\bibinfo  {journal} {arXiv}\ ,\ \bibinfo {pages}
  {1807.05795}} (\bibinfo {year} {2018})}\BibitemShut {NoStop}%
\bibitem [{\citenamefont {Lukin}\ \emph {et~al.}(2001)\citenamefont {Lukin},
  \citenamefont {Fleischhauer}, \citenamefont {Cote}, \citenamefont {Duan},
  \citenamefont {Jaksch}, \citenamefont {Cirac},\ and\ \citenamefont
  {Zoller}}]{Lukin01}%
  \BibitemOpen
  \bibfield  {author} {\bibinfo {author} {\bibfnamefont {M.}~\bibnamefont
  {Lukin}}, \bibinfo {author} {\bibfnamefont {M.}~\bibnamefont {Fleischhauer}},
  \bibinfo {author} {\bibfnamefont {R.}~\bibnamefont {Cote}}, \bibinfo {author}
  {\bibfnamefont {L.}~\bibnamefont {Duan}}, \bibinfo {author} {\bibfnamefont
  {D.}~\bibnamefont {Jaksch}}, \bibinfo {author} {\bibfnamefont
  {J.}~\bibnamefont {Cirac}}, \ and\ \bibinfo {author} {\bibfnamefont
  {P.}~\bibnamefont {Zoller}},\ }\href@noop {} {\bibfield  {journal} {\bibinfo
  {journal} {Physical Review Letters}\ }\textbf {\bibinfo {volume} {87}},\
  \bibinfo {pages} {037901} (\bibinfo {year} {2001})}\BibitemShut {NoStop}%
\bibitem [{\citenamefont {Ebert}\ \emph {et~al.}(2015)\citenamefont {Ebert},
  \citenamefont {Kwon}, \citenamefont {Walker},\ and\ \citenamefont
  {Saffman}}]{Ebert15}%
  \BibitemOpen
  \bibfield  {author} {\bibinfo {author} {\bibfnamefont {M.}~\bibnamefont
  {Ebert}}, \bibinfo {author} {\bibfnamefont {M.}~\bibnamefont {Kwon}},
  \bibinfo {author} {\bibfnamefont {T.~G.}\ \bibnamefont {Walker}}, \ and\
  \bibinfo {author} {\bibfnamefont {M.}~\bibnamefont {Saffman}},\ }\href@noop
  {} {\bibfield  {journal} {\bibinfo  {journal} {Physical Review Letters}\
  }\textbf {\bibinfo {volume} {115}},\ \bibinfo {pages} {093601} (\bibinfo
  {year} {2015})}\BibitemShut {NoStop}%
\bibitem [{\citenamefont {Balewski}\ \emph {et~al.}(2013)\citenamefont
  {Balewski}, \citenamefont {Krupp}, \citenamefont {Gaj}, \citenamefont
  {Peter}, \citenamefont {B\"uchler}, \citenamefont {L\"ow}, \citenamefont
  {Hofferberth},\ and\ \citenamefont {Pfau}}]{Balewski13}%
  \BibitemOpen
  \bibfield  {author} {\bibinfo {author} {\bibfnamefont {J.~B.}\ \bibnamefont
  {Balewski}}, \bibinfo {author} {\bibfnamefont {A.~T.}\ \bibnamefont {Krupp}},
  \bibinfo {author} {\bibfnamefont {A.}~\bibnamefont {Gaj}}, \bibinfo {author}
  {\bibfnamefont {D.}~\bibnamefont {Peter}}, \bibinfo {author} {\bibfnamefont
  {H.~P.}\ \bibnamefont {B\"uchler}}, \bibinfo {author} {\bibfnamefont
  {R.}~\bibnamefont {L\"ow}}, \bibinfo {author} {\bibfnamefont
  {S.}~\bibnamefont {Hofferberth}}, \ and\ \bibinfo {author} {\bibfnamefont
  {T.}~\bibnamefont {Pfau}},\ }\href@noop {} {\bibfield  {journal} {\bibinfo
  {journal} {Nature}\ }\textbf {\bibinfo {volume} {502}},\ \bibinfo {pages}
  {664} (\bibinfo {year} {2013})}\BibitemShut {NoStop}%
\bibitem [{\citenamefont {Barredo}\ \emph {et~al.}(2015)\citenamefont
  {Barredo}, \citenamefont {Labuhn}, \citenamefont {Ravets}, \citenamefont
  {Lahaye}, \citenamefont {Browaeys},\ and\ \citenamefont {Adams}}]{Barredo14}%
  \BibitemOpen
  \bibfield  {author} {\bibinfo {author} {\bibfnamefont {D.}~\bibnamefont
  {Barredo}}, \bibinfo {author} {\bibfnamefont {H.}~\bibnamefont {Labuhn}},
  \bibinfo {author} {\bibfnamefont {S.}~\bibnamefont {Ravets}}, \bibinfo
  {author} {\bibfnamefont {T.}~\bibnamefont {Lahaye}}, \bibinfo {author}
  {\bibfnamefont {A.}~\bibnamefont {Browaeys}}, \ and\ \bibinfo {author}
  {\bibfnamefont {C.~S.}\ \bibnamefont {Adams}},\ }\href@noop {} {\bibfield
  {journal} {\bibinfo  {journal} {Physical Review Letters}\ }\textbf {\bibinfo
  {volume} {114}},\ \bibinfo {pages} {113002} (\bibinfo {year}
  {2015})}\BibitemShut {NoStop}%
\bibitem [{\citenamefont {Bernien}\ \emph {et~al.}(2017)\citenamefont
  {Bernien}, \citenamefont {Schwartz}, \citenamefont {Keesling}, \citenamefont
  {Levine}, \citenamefont {Omran}, \citenamefont {Pichler}, \citenamefont
  {Choi}, \citenamefont {Zibrov}, \citenamefont {Endres}, \citenamefont
  {Greiner}, \citenamefont {Vuleti\'c},\ and\ \citenamefont
  {Lukin}}]{Bernien17}%
  \BibitemOpen
  \bibfield  {author} {\bibinfo {author} {\bibfnamefont {H.}~\bibnamefont
  {Bernien}}, \bibinfo {author} {\bibfnamefont {S.}~\bibnamefont {Schwartz}},
  \bibinfo {author} {\bibfnamefont {A.}~\bibnamefont {Keesling}}, \bibinfo
  {author} {\bibfnamefont {H.}~\bibnamefont {Levine}}, \bibinfo {author}
  {\bibfnamefont {A.}~\bibnamefont {Omran}}, \bibinfo {author} {\bibfnamefont
  {H.}~\bibnamefont {Pichler}}, \bibinfo {author} {\bibfnamefont
  {S.}~\bibnamefont {Choi}}, \bibinfo {author} {\bibfnamefont {A.~S.}\
  \bibnamefont {Zibrov}}, \bibinfo {author} {\bibfnamefont {M.}~\bibnamefont
  {Endres}}, \bibinfo {author} {\bibfnamefont {M.}~\bibnamefont {Greiner}},
  \bibinfo {author} {\bibfnamefont {V.}~\bibnamefont {Vuleti\'c}}, \ and\
  \bibinfo {author} {\bibfnamefont {M.~D.}\ \bibnamefont {Lukin}},\ }\href@noop
  {} {\bibfield  {journal} {\bibinfo  {journal} {Nature}\ }\textbf {\bibinfo
  {volume} {551}},\ \bibinfo {pages} {579} (\bibinfo {year}
  {2017})}\BibitemShut {NoStop}%
\bibitem [{\citenamefont {Zeiher}\ \emph {et~al.}(2016)\citenamefont {Zeiher},
  \citenamefont {Bijnen}, \citenamefont {Schauß}, \citenamefont {Hild},
  \citenamefont {Choi}, \citenamefont {Pohl}, \citenamefont {Bloch},\ and\
  \citenamefont {Gross}}]{Zeiher16}%
  \BibitemOpen
  \bibfield  {author} {\bibinfo {author} {\bibfnamefont {J.}~\bibnamefont
  {Zeiher}}, \bibinfo {author} {\bibfnamefont {R.~v.}\ \bibnamefont {Bijnen}},
  \bibinfo {author} {\bibfnamefont {P.}~\bibnamefont {Schauß}}, \bibinfo
  {author} {\bibfnamefont {S.}~\bibnamefont {Hild}}, \bibinfo {author}
  {\bibfnamefont {J.-y.}\ \bibnamefont {Choi}}, \bibinfo {author}
  {\bibfnamefont {T.}~\bibnamefont {Pohl}}, \bibinfo {author} {\bibfnamefont
  {I.}~\bibnamefont {Bloch}}, \ and\ \bibinfo {author} {\bibfnamefont
  {C.}~\bibnamefont {Gross}},\ }\href@noop {} {\bibfield  {journal} {\bibinfo
  {journal} {Nature Physics}\ }\textbf {\bibinfo {volume} {12}},\ \bibinfo
  {pages} {1095} (\bibinfo {year} {2016})}\BibitemShut {NoStop}%
\bibitem [{\citenamefont {L\"ow}\ \emph {et~al.}(2012)\citenamefont {L\"ow},
  \citenamefont {Weimer}, \citenamefont {Nipper}, \citenamefont {Balewski},
  \citenamefont {Butscher}, \citenamefont {Büchler},\ and\ \citenamefont
  {Pfau}}]{Low12}%
  \BibitemOpen
  \bibfield  {author} {\bibinfo {author} {\bibfnamefont {R.}~\bibnamefont
  {L\"ow}}, \bibinfo {author} {\bibfnamefont {H.}~\bibnamefont {Weimer}},
  \bibinfo {author} {\bibfnamefont {J.}~\bibnamefont {Nipper}}, \bibinfo
  {author} {\bibfnamefont {J.~B.}\ \bibnamefont {Balewski}}, \bibinfo {author}
  {\bibfnamefont {B.}~\bibnamefont {Butscher}}, \bibinfo {author}
  {\bibfnamefont {H.~P.}\ \bibnamefont {Büchler}}, \ and\ \bibinfo {author}
  {\bibfnamefont {T.}~\bibnamefont {Pfau}},\ }\href@noop {} {\bibfield
  {journal} {\bibinfo  {journal} {Journal of Physics B: Atomic, Molecular and
  Optical Physics}\ }\textbf {\bibinfo {volume} {45}},\ \bibinfo {pages}
  {113001} (\bibinfo {year} {2012})}\BibitemShut {NoStop}%
\bibitem [{\citenamefont {Goldschmidt}\ \emph {et~al.}(2016)\citenamefont
  {Goldschmidt}, \citenamefont {Boulier}, \citenamefont {Brown}, \citenamefont
  {Koller}, \citenamefont {Young}, \citenamefont {Gorshkov}, \citenamefont
  {Rolston},\ and\ \citenamefont {Porto}}]{Goldschmidt16}%
  \BibitemOpen
  \bibfield  {author} {\bibinfo {author} {\bibfnamefont {E.~A.}\ \bibnamefont
  {Goldschmidt}}, \bibinfo {author} {\bibfnamefont {T.}~\bibnamefont
  {Boulier}}, \bibinfo {author} {\bibfnamefont {R.~C.}\ \bibnamefont {Brown}},
  \bibinfo {author} {\bibfnamefont {S.~B.}\ \bibnamefont {Koller}}, \bibinfo
  {author} {\bibfnamefont {J.~T.}\ \bibnamefont {Young}}, \bibinfo {author}
  {\bibfnamefont {A.~V.}\ \bibnamefont {Gorshkov}}, \bibinfo {author}
  {\bibfnamefont {S.~L.}\ \bibnamefont {Rolston}}, \ and\ \bibinfo {author}
  {\bibfnamefont {J.~V.}\ \bibnamefont {Porto}},\ }\href@noop {} {\bibfield
  {journal} {\bibinfo  {journal} {Physical Review Letters}\ }\textbf {\bibinfo
  {volume} {116}},\ \bibinfo {pages} {113001} (\bibinfo {year}
  {2016})}\BibitemShut {NoStop}%
\bibitem [{\citenamefont {Young}\ \emph {et~al.}(2018)\citenamefont {Young},
  \citenamefont {Boulier}, \citenamefont {Magnan}, \citenamefont {Goldschmidt},
  \citenamefont {Wilson}, \citenamefont {Rolston}, \citenamefont {Porto},\ and\
  \citenamefont {Gorshkov}}]{Young17}%
  \BibitemOpen
  \bibfield  {author} {\bibinfo {author} {\bibfnamefont {J.~T.}\ \bibnamefont
  {Young}}, \bibinfo {author} {\bibfnamefont {T.}~\bibnamefont {Boulier}},
  \bibinfo {author} {\bibfnamefont {E.}~\bibnamefont {Magnan}}, \bibinfo
  {author} {\bibfnamefont {E.~A.}\ \bibnamefont {Goldschmidt}}, \bibinfo
  {author} {\bibfnamefont {R.~M.}\ \bibnamefont {Wilson}}, \bibinfo {author}
  {\bibfnamefont {S.~L.}\ \bibnamefont {Rolston}}, \bibinfo {author}
  {\bibfnamefont {J.~V.}\ \bibnamefont {Porto}}, \ and\ \bibinfo {author}
  {\bibfnamefont {A.~V.}\ \bibnamefont {Gorshkov}},\ }\href@noop {} {\bibfield
  {journal} {\bibinfo  {journal} {Physical Review A}\ }\textbf {\bibinfo
  {volume} {97}},\ \bibinfo {pages} {023424} (\bibinfo {year}
  {2018})}\BibitemShut {NoStop}%
\bibitem [{\citenamefont {Boulier}\ \emph {et~al.}(2017)\citenamefont
  {Boulier}, \citenamefont {Magnan}, \citenamefont {Bracamontes}, \citenamefont
  {Maslek}, \citenamefont {Goldschmidt}, \citenamefont {Young}, \citenamefont
  {Gorshkov}, \citenamefont {Rolston},\ and\ \citenamefont
  {Porto}}]{Boulier17}%
  \BibitemOpen
  \bibfield  {author} {\bibinfo {author} {\bibfnamefont {T.}~\bibnamefont
  {Boulier}}, \bibinfo {author} {\bibfnamefont {E.}~\bibnamefont {Magnan}},
  \bibinfo {author} {\bibfnamefont {C.}~\bibnamefont {Bracamontes}}, \bibinfo
  {author} {\bibfnamefont {J.}~\bibnamefont {Maslek}}, \bibinfo {author}
  {\bibfnamefont {E.~A.}\ \bibnamefont {Goldschmidt}}, \bibinfo {author}
  {\bibfnamefont {J.~T.}\ \bibnamefont {Young}}, \bibinfo {author}
  {\bibfnamefont {A.~V.}\ \bibnamefont {Gorshkov}}, \bibinfo {author}
  {\bibfnamefont {S.~L.}\ \bibnamefont {Rolston}}, \ and\ \bibinfo {author}
  {\bibfnamefont {J.~V.}\ \bibnamefont {Porto}},\ }\href@noop {} {\bibfield
  {journal} {\bibinfo  {journal} {Physical Review A}\ }\textbf {\bibinfo
  {volume} {96}},\ \bibinfo {pages} {053409} (\bibinfo {year}
  {2017})}\BibitemShut {NoStop}%
\bibitem [{\citenamefont {Dudin}\ \emph {et~al.}(2012)\citenamefont {Dudin},
  \citenamefont {Li}, \citenamefont {Bariani},\ and\ \citenamefont
  {Kuzmich}}]{Dudin12}%
  \BibitemOpen
  \bibfield  {author} {\bibinfo {author} {\bibfnamefont {Y.~O.}\ \bibnamefont
  {Dudin}}, \bibinfo {author} {\bibfnamefont {L.}~\bibnamefont {Li}}, \bibinfo
  {author} {\bibfnamefont {F.}~\bibnamefont {Bariani}}, \ and\ \bibinfo
  {author} {\bibfnamefont {A.}~\bibnamefont {Kuzmich}},\ }\href@noop {}
  {\bibfield  {journal} {\bibinfo  {journal} {Nature Physics}\ }\textbf
  {\bibinfo {volume} {8}},\ \bibinfo {pages} {790} (\bibinfo {year}
  {2012})}\BibitemShut {NoStop}%
\bibitem [{\citenamefont {Helmrich}\ \emph {et~al.}(2018)\citenamefont
  {Helmrich}, \citenamefont {Arias},\ and\ \citenamefont
  {Whitlock}}]{Whitlock16}%
  \BibitemOpen
  \bibfield  {author} {\bibinfo {author} {\bibfnamefont {S.}~\bibnamefont
  {Helmrich}}, \bibinfo {author} {\bibfnamefont {A.}~\bibnamefont {Arias}}, \
  and\ \bibinfo {author} {\bibfnamefont {S.}~\bibnamefont {Whitlock}},\
  }\href@noop {} {\bibfield  {journal} {\bibinfo  {journal} {Physical Review
  A}\ }\textbf {\bibinfo {volume} {98}},\ \bibinfo {pages} {022109} (\bibinfo
  {year} {2018})}\BibitemShut {NoStop}%
\bibitem [{\citenamefont {Viteau}\ \emph {et~al.}(2011)\citenamefont {Viteau},
  \citenamefont {Bason}, \citenamefont {Radogostowicz}, \citenamefont
  {Malossi}, \citenamefont {Ciampini}, \citenamefont {Morsch},\ and\
  \citenamefont {Arimondo}}]{Viteau11}%
  \BibitemOpen
  \bibfield  {author} {\bibinfo {author} {\bibfnamefont {M.}~\bibnamefont
  {Viteau}}, \bibinfo {author} {\bibfnamefont {M.~G.}\ \bibnamefont {Bason}},
  \bibinfo {author} {\bibfnamefont {J.}~\bibnamefont {Radogostowicz}}, \bibinfo
  {author} {\bibfnamefont {N.}~\bibnamefont {Malossi}}, \bibinfo {author}
  {\bibfnamefont {D.}~\bibnamefont {Ciampini}}, \bibinfo {author}
  {\bibfnamefont {O.}~\bibnamefont {Morsch}}, \ and\ \bibinfo {author}
  {\bibfnamefont {E.}~\bibnamefont {Arimondo}},\ }\href {\doibase
  10.1103/PhysRevLett.107.060402} {\bibfield  {journal} {\bibinfo  {journal}
  {Phys. Rev. Lett.}\ }\textbf {\bibinfo {volume} {107}},\ \bibinfo {pages}
  {060402} (\bibinfo {year} {2011})}\BibitemShut {NoStop}%
\bibitem [{\citenamefont {Zeiher}\ \emph {et~al.}(2015)\citenamefont {Zeiher},
  \citenamefont {Schauss}, \citenamefont {Hild}, \citenamefont {Macr\`i},
  \citenamefont {Bloch},\ and\ \citenamefont {Gross}}]{Zeiher15}%
  \BibitemOpen
  \bibfield  {author} {\bibinfo {author} {\bibfnamefont {J.}~\bibnamefont
  {Zeiher}}, \bibinfo {author} {\bibfnamefont {P.}~\bibnamefont {Schauss}},
  \bibinfo {author} {\bibfnamefont {S.}~\bibnamefont {Hild}}, \bibinfo {author}
  {\bibfnamefont {T.}~\bibnamefont {Macr\`i}}, \bibinfo {author} {\bibfnamefont
  {I.}~\bibnamefont {Bloch}}, \ and\ \bibinfo {author} {\bibfnamefont
  {C.}~\bibnamefont {Gross}},\ }\href@noop {} {\bibfield  {journal} {\bibinfo
  {journal} {Physical Review X}\ }\textbf {\bibinfo {volume} {5}},\ \bibinfo
  {pages} {031015} (\bibinfo {year} {2015})}\BibitemShut {NoStop}%
\bibitem [{\citenamefont {Barredo}\ \emph {et~al.}(2018)\citenamefont
  {Barredo}, \citenamefont {Lienhard}, \citenamefont {de~L\'{e}s\'e{}leuc},
  \citenamefont {Lahaye},\ and\ \citenamefont {Browaeys}}]{Barredo18}%
  \BibitemOpen
  \bibfield  {author} {\bibinfo {author} {\bibfnamefont {D.}~\bibnamefont
  {Barredo}}, \bibinfo {author} {\bibfnamefont {V.}~\bibnamefont {Lienhard}},
  \bibinfo {author} {\bibfnamefont {S.}~\bibnamefont {de~L\'{e}s\'e{}leuc}},
  \bibinfo {author} {\bibfnamefont {T.}~\bibnamefont {Lahaye}}, \ and\ \bibinfo
  {author} {\bibfnamefont {A.}~\bibnamefont {Browaeys}},\ }\href@noop {}
  {\bibfield  {journal} {\bibinfo  {journal} {Nature}\ }\textbf {\bibinfo
  {volume} {561}} (\bibinfo {year} {2018})}\BibitemShut {NoStop}%
\bibitem [{\citenamefont {Reichel}\ and\ \citenamefont
  {Vuleti{\'c}}(2011)}]{Reichel11}%
  \BibitemOpen
  \bibinfo {editor} {\bibfnamefont {J.}~\bibnamefont {Reichel}}\ and\ \bibinfo
  {editor} {\bibfnamefont {V.}~\bibnamefont {Vuleti{\'c}}},\ eds.,\ \href@noop
  {} {\emph {\bibinfo {title} {Atom Chips}}}\ (\bibinfo  {publisher}
  {Wiley-VCH},\ \bibinfo {year} {2011})\BibitemShut {NoStop}%
\bibitem [{\citenamefont {Petrosyan}\ \emph {et~al.}(2009)\citenamefont
  {Petrosyan}, \citenamefont {Bensky}, \citenamefont {Kurizki}, \citenamefont
  {Mazets}, \citenamefont {Majer},\ and\ \citenamefont
  {Schmiedmayer}}]{Petrosyan09}%
  \BibitemOpen
  \bibfield  {author} {\bibinfo {author} {\bibfnamefont {D.}~\bibnamefont
  {Petrosyan}}, \bibinfo {author} {\bibfnamefont {G.}~\bibnamefont {Bensky}},
  \bibinfo {author} {\bibfnamefont {G.}~\bibnamefont {Kurizki}}, \bibinfo
  {author} {\bibfnamefont {I.}~\bibnamefont {Mazets}}, \bibinfo {author}
  {\bibfnamefont {J.}~\bibnamefont {Majer}}, \ and\ \bibinfo {author}
  {\bibfnamefont {J.}~\bibnamefont {Schmiedmayer}},\ }\href {\doibase
  10.1103/PhysRevA.79.040304} {\bibfield  {journal} {\bibinfo  {journal} {Phys.
  Rev. A}\ }\textbf {\bibinfo {volume} {79}},\ \bibinfo {pages} {040304}
  (\bibinfo {year} {2009})}\BibitemShut {NoStop}%
\bibitem [{\citenamefont {Xiang}\ \emph {et~al.}(2013)\citenamefont {Xiang},
  \citenamefont {Ashhab}, \citenamefont {You},\ and\ \citenamefont
  {Nori}}]{Xiang13}%
  \BibitemOpen
  \bibfield  {author} {\bibinfo {author} {\bibfnamefont {Z.-L.}\ \bibnamefont
  {Xiang}}, \bibinfo {author} {\bibfnamefont {S.}~\bibnamefont {Ashhab}},
  \bibinfo {author} {\bibfnamefont {J.~Q.}\ \bibnamefont {You}}, \ and\
  \bibinfo {author} {\bibfnamefont {F.}~\bibnamefont {Nori}},\ }\href {\doibase
  10.1103/RevModPhys.85.623} {\bibfield  {journal} {\bibinfo  {journal} {Rev.
  Mod. Phys.}\ }\textbf {\bibinfo {volume} {85}},\ \bibinfo {pages} {623}
  (\bibinfo {year} {2013})}\BibitemShut {NoStop}%
\bibitem [{\citenamefont {Tauschinsky}\ \emph {et~al.}(2010)\citenamefont
  {Tauschinsky}, \citenamefont {Thijssen}, \citenamefont {Whitlock},
  \citenamefont {van Linden van~den Heuvell},\ and\ \citenamefont
  {Spreeuw}}]{Tauschinsky10}%
  \BibitemOpen
  \bibfield  {author} {\bibinfo {author} {\bibfnamefont {A.}~\bibnamefont
  {Tauschinsky}}, \bibinfo {author} {\bibfnamefont {R.~M.~T.}\ \bibnamefont
  {Thijssen}}, \bibinfo {author} {\bibfnamefont {S.}~\bibnamefont {Whitlock}},
  \bibinfo {author} {\bibfnamefont {H.~B.}\ \bibnamefont {van Linden van~den
  Heuvell}}, \ and\ \bibinfo {author} {\bibfnamefont {R.~J.~C.}\ \bibnamefont
  {Spreeuw}},\ }\href@noop {} {\bibfield  {journal} {\bibinfo  {journal}
  {Physical Review A}\ }\textbf {\bibinfo {volume} {81}},\ \bibinfo {pages}
  {063411} (\bibinfo {year} {2010})}\BibitemShut {NoStop}%
\bibitem [{\citenamefont {Hermann-Avigliano}\ \emph {et~al.}(2014)\citenamefont
  {Hermann-Avigliano}, \citenamefont {{Celistrino Teixeira}}, \citenamefont
  {Nguyen}, \citenamefont {Cantat-Moltrecht}, \citenamefont {Nogues},
  \citenamefont {Dotsenko}, \citenamefont {Gleyzes}, \citenamefont {Raimond},
  \citenamefont {Haroche},\ and\ \citenamefont {Brune}}]{Hermann14}%
  \BibitemOpen
  \bibfield  {author} {\bibinfo {author} {\bibfnamefont {C.}~\bibnamefont
  {Hermann-Avigliano}}, \bibinfo {author} {\bibfnamefont {R.}~\bibnamefont
  {{Celistrino Teixeira}}}, \bibinfo {author} {\bibfnamefont {T.~L.}\
  \bibnamefont {Nguyen}}, \bibinfo {author} {\bibfnamefont {T.}~\bibnamefont
  {Cantat-Moltrecht}}, \bibinfo {author} {\bibfnamefont {G.}~\bibnamefont
  {Nogues}}, \bibinfo {author} {\bibfnamefont {I.}~\bibnamefont {Dotsenko}},
  \bibinfo {author} {\bibfnamefont {S.}~\bibnamefont {Gleyzes}}, \bibinfo
  {author} {\bibfnamefont {J.~M.}\ \bibnamefont {Raimond}}, \bibinfo {author}
  {\bibfnamefont {S.}~\bibnamefont {Haroche}}, \ and\ \bibinfo {author}
  {\bibfnamefont {M.}~\bibnamefont {Brune}},\ }\href@noop {} {\bibfield
  {journal} {\bibinfo  {journal} {Physical Review A}\ }\textbf {\bibinfo
  {volume} {90}},\ \bibinfo {pages} {040502(R)} (\bibinfo {year}
  {2014})}\BibitemShut {NoStop}%
\bibitem [{\citenamefont {Cisternas}\ \emph {et~al.}(2017)\citenamefont
  {Cisternas}, \citenamefont {de~Hond}, \citenamefont {Lochead}, \citenamefont
  {Spreeuw}, \citenamefont {van Linden van~den Heuvell},\ and\ \citenamefont
  {van Druten}}]{Cisternas17}%
  \BibitemOpen
  \bibfield  {author} {\bibinfo {author} {\bibfnamefont {N.}~\bibnamefont
  {Cisternas}}, \bibinfo {author} {\bibfnamefont {J.}~\bibnamefont {de~Hond}},
  \bibinfo {author} {\bibfnamefont {G.}~\bibnamefont {Lochead}}, \bibinfo
  {author} {\bibfnamefont {R.~J.~C.}\ \bibnamefont {Spreeuw}}, \bibinfo
  {author} {\bibfnamefont {H.~B.}\ \bibnamefont {van Linden van~den Heuvell}},
  \ and\ \bibinfo {author} {\bibfnamefont {N.~J.}\ \bibnamefont {van Druten}},\
  }\href@noop {} {\bibfield  {journal} {\bibinfo  {journal} {Physical Review
  A}\ }\textbf {\bibinfo {volume} {96}},\ \bibinfo {pages} {013425} (\bibinfo
  {year} {2017})}\BibitemShut {NoStop}%
\bibitem [{\citenamefont {Davtyan}\ \emph {et~al.}(2018)\citenamefont
  {Davtyan}, \citenamefont {Machluf}, \citenamefont {Soudijn}, \citenamefont
  {Naber}, \citenamefont {van Druten}, \citenamefont {van Linden van~den
  Heuvell},\ and\ \citenamefont {Spreeuw}}]{Davtyan18}%
  \BibitemOpen
  \bibfield  {author} {\bibinfo {author} {\bibfnamefont {D.}~\bibnamefont
  {Davtyan}}, \bibinfo {author} {\bibfnamefont {S.}~\bibnamefont {Machluf}},
  \bibinfo {author} {\bibfnamefont {M.~L.}\ \bibnamefont {Soudijn}}, \bibinfo
  {author} {\bibfnamefont {J.~B.}\ \bibnamefont {Naber}}, \bibinfo {author}
  {\bibfnamefont {N.~J.}\ \bibnamefont {van Druten}}, \bibinfo {author}
  {\bibfnamefont {H.~B.}\ \bibnamefont {van Linden van~den Heuvell}}, \ and\
  \bibinfo {author} {\bibfnamefont {R.~J.~C.}\ \bibnamefont {Spreeuw}},\
  }\href@noop {} {\bibfield  {journal} {\bibinfo  {journal} {Physical Review
  A}\ }\textbf {\bibinfo {volume} {97}},\ \bibinfo {pages} {023418} (\bibinfo
  {year} {2018})}\BibitemShut {NoStop}%
\bibitem [{\citenamefont {Hogan}\ \emph {et~al.}(2012)\citenamefont {Hogan},
  \citenamefont {Agner}, \citenamefont {Merkt}, \citenamefont {Thiele},
  \citenamefont {Filipp},\ and\ \citenamefont {Wallraff}}]{Hogan12}%
  \BibitemOpen
  \bibfield  {author} {\bibinfo {author} {\bibfnamefont {S.~D.}\ \bibnamefont
  {Hogan}}, \bibinfo {author} {\bibfnamefont {J.~A.}\ \bibnamefont {Agner}},
  \bibinfo {author} {\bibfnamefont {F.}~\bibnamefont {Merkt}}, \bibinfo
  {author} {\bibfnamefont {T.}~\bibnamefont {Thiele}}, \bibinfo {author}
  {\bibfnamefont {S.}~\bibnamefont {Filipp}}, \ and\ \bibinfo {author}
  {\bibfnamefont {A.}~\bibnamefont {Wallraff}},\ }\href {\doibase
  10.1103/PhysRevLett.108.063004} {\bibfield  {journal} {\bibinfo  {journal}
  {Phys. Rev. Lett.}\ }\textbf {\bibinfo {volume} {108}},\ \bibinfo {pages}
  {063004} (\bibinfo {year} {2012})}\BibitemShut {NoStop}%
\bibitem [{\citenamefont {Thiele}\ \emph {et~al.}(2015)\citenamefont {Thiele},
  \citenamefont {Deiglmayr}, \citenamefont {Stammeier}, \citenamefont {Agner},
  \citenamefont {Schmutz}, \citenamefont {Merkt},\ and\ \citenamefont
  {Wallraff}}]{Thiele15}%
  \BibitemOpen
  \bibfield  {author} {\bibinfo {author} {\bibfnamefont {T.}~\bibnamefont
  {Thiele}}, \bibinfo {author} {\bibfnamefont {J.}~\bibnamefont {Deiglmayr}},
  \bibinfo {author} {\bibfnamefont {M.}~\bibnamefont {Stammeier}}, \bibinfo
  {author} {\bibfnamefont {J.-A.}\ \bibnamefont {Agner}}, \bibinfo {author}
  {\bibfnamefont {H.}~\bibnamefont {Schmutz}}, \bibinfo {author} {\bibfnamefont
  {F.}~\bibnamefont {Merkt}}, \ and\ \bibinfo {author} {\bibfnamefont
  {A.}~\bibnamefont {Wallraff}},\ }\href {\doibase 10.1103/PhysRevA.92.063425}
  {\bibfield  {journal} {\bibinfo  {journal} {Phys. Rev. A}\ }\textbf {\bibinfo
  {volume} {92}},\ \bibinfo {pages} {063425} (\bibinfo {year}
  {2015})}\BibitemShut {NoStop}%
\bibitem [{\citenamefont {Celistrino~Teixeira}\ \emph
  {et~al.}(2015)\citenamefont {Celistrino~Teixeira}, \citenamefont
  {Hermann-Avigliano}, \citenamefont {Nguyen}, \citenamefont
  {Cantat-Moltrecht}, \citenamefont {Raimond}, \citenamefont {Haroche},
  \citenamefont {Gleyzes},\ and\ \citenamefont {Brune}}]{Teixeira15}%
  \BibitemOpen
  \bibfield  {author} {\bibinfo {author} {\bibfnamefont {R.}~\bibnamefont
  {Celistrino~Teixeira}}, \bibinfo {author} {\bibfnamefont {C.}~\bibnamefont
  {Hermann-Avigliano}}, \bibinfo {author} {\bibfnamefont {T.~L.}\ \bibnamefont
  {Nguyen}}, \bibinfo {author} {\bibfnamefont {T.}~\bibnamefont
  {Cantat-Moltrecht}}, \bibinfo {author} {\bibfnamefont {J.}~\bibnamefont
  {Raimond}}, \bibinfo {author} {\bibfnamefont {S.}~\bibnamefont {Haroche}},
  \bibinfo {author} {\bibfnamefont {S.}~\bibnamefont {Gleyzes}}, \ and\
  \bibinfo {author} {\bibfnamefont {M.}~\bibnamefont {Brune}},\ }\href@noop {}
  {\bibfield  {journal} {\bibinfo  {journal} {Physical Review Letters}\
  }\textbf {\bibinfo {volume} {115}},\ \bibinfo {pages} {013001} (\bibinfo
  {year} {2015})}\BibitemShut {NoStop}%
\bibitem [{\citenamefont {{van Es}}\ \emph {et~al.}(2010)\citenamefont {{van
  Es}}, \citenamefont {Wicke}, \citenamefont {{van Amerongen}}, \citenamefont
  {R\'etif}, \citenamefont {Whitlock},\ and\ \citenamefont {{van
  Druten}}}]{vanEs10}%
  \BibitemOpen
  \bibfield  {author} {\bibinfo {author} {\bibfnamefont {J.~J.~P.}\
  \bibnamefont {{van Es}}}, \bibinfo {author} {\bibfnamefont {P.}~\bibnamefont
  {Wicke}}, \bibinfo {author} {\bibfnamefont {A.~H.}\ \bibnamefont {{van
  Amerongen}}}, \bibinfo {author} {\bibfnamefont {C.}~\bibnamefont {R\'etif}},
  \bibinfo {author} {\bibfnamefont {S.}~\bibnamefont {Whitlock}}, \ and\
  \bibinfo {author} {\bibfnamefont {N.~J.}\ \bibnamefont {{van Druten}}},\
  }\href@noop {} {\bibfield  {journal} {\bibinfo  {journal} {Journal of Physics
  B: Atomic, Molecular and Optical Physics}\ }\textbf {\bibinfo {volume}
  {43}},\ \bibinfo {pages} {155002} (\bibinfo {year} {2010})}\BibitemShut
  {NoStop}%
\bibitem [{\citenamefont {Brion}\ \emph {et~al.}(2007)\citenamefont {Brion},
  \citenamefont {Pedersen},\ and\ \citenamefont {M{\o}lmer}}]{Brion07}%
  \BibitemOpen
  \bibfield  {author} {\bibinfo {author} {\bibfnamefont {E.}~\bibnamefont
  {Brion}}, \bibinfo {author} {\bibfnamefont {L.~H.}\ \bibnamefont {Pedersen}},
  \ and\ \bibinfo {author} {\bibfnamefont {K.}~\bibnamefont {M{\o}lmer}},\
  }\href@noop {} {\bibfield  {journal} {\bibinfo  {journal} {Journal of Physics
  A: Mathematical and Theoretical}\ }\textbf {\bibinfo {volume} {40}},\
  \bibinfo {pages} {1033} (\bibinfo {year} {2007})}\BibitemShut {NoStop}%
\bibitem [{\citenamefont {de~Hond}\ \emph {et~al.}(2017)\citenamefont
  {de~Hond}, \citenamefont {Cisternas}, \citenamefont {Lochead},\ and\
  \citenamefont {van Druten}}]{deHond17}%
  \BibitemOpen
  \bibfield  {author} {\bibinfo {author} {\bibfnamefont {J.}~\bibnamefont
  {de~Hond}}, \bibinfo {author} {\bibfnamefont {N.}~\bibnamefont {Cisternas}},
  \bibinfo {author} {\bibfnamefont {G.}~\bibnamefont {Lochead}}, \ and\
  \bibinfo {author} {\bibfnamefont {N.~J.}\ \bibnamefont {van Druten}},\
  }\href@noop {} {\bibfield  {journal} {\bibinfo  {journal} {Applied Optics}\
  }\textbf {\bibinfo {volume} {56}},\ \bibinfo {pages} {5436} (\bibinfo {year}
  {2017})}\BibitemShut {NoStop}%
\bibitem [{\citenamefont {Gallagher}(1994)}]{Gallagher94}%
  \BibitemOpen
  \bibfield  {author} {\bibinfo {author} {\bibfnamefont {T.~F.}\ \bibnamefont
  {Gallagher}},\ }\href@noop {} {\emph {\bibinfo {title} {{Rydberg} atoms}}}\
  (\bibinfo  {publisher} {Cambridge University Press},\ \bibinfo {year}
  {1994})\BibitemShut {NoStop}%
\bibitem [{\citenamefont {Šibalić}\ \emph {et~al.}(2017)\citenamefont
  {Šibalić}, \citenamefont {Pritchard}, \citenamefont {Adams},\ and\
  \citenamefont {Weatherill}}]{Sibalic17}%
  \BibitemOpen
  \bibfield  {author} {\bibinfo {author} {\bibfnamefont {N.}~\bibnamefont
  {Šibalić}}, \bibinfo {author} {\bibfnamefont {J.~D.}\ \bibnamefont
  {Pritchard}}, \bibinfo {author} {\bibfnamefont {C.~S.}\ \bibnamefont
  {Adams}}, \ and\ \bibinfo {author} {\bibfnamefont {K.~J.}\ \bibnamefont
  {Weatherill}},\ }\href@noop {} {\bibfield  {journal} {\bibinfo  {journal}
  {Computer Physics Communications}\ }\textbf {\bibinfo {volume} {220}},\
  \bibinfo {pages} {319} (\bibinfo {year} {2017})}\BibitemShut {NoStop}%
\bibitem [{\citenamefont {Weber}\ \emph {et~al.}(2017)\citenamefont {Weber},
  \citenamefont {Tresp}, \citenamefont {Menke}, \citenamefont {Urvoy},
  \citenamefont {Firstenberg}, \citenamefont {B\"uchler},\ and\ \citenamefont
  {Hofferberth}}]{Weber17}%
  \BibitemOpen
  \bibfield  {author} {\bibinfo {author} {\bibfnamefont {S.}~\bibnamefont
  {Weber}}, \bibinfo {author} {\bibfnamefont {C.}~\bibnamefont {Tresp}},
  \bibinfo {author} {\bibfnamefont {H.}~\bibnamefont {Menke}}, \bibinfo
  {author} {\bibfnamefont {A.}~\bibnamefont {Urvoy}}, \bibinfo {author}
  {\bibfnamefont {O.}~\bibnamefont {Firstenberg}}, \bibinfo {author}
  {\bibfnamefont {H.~P.}\ \bibnamefont {B\"uchler}}, \ and\ \bibinfo {author}
  {\bibfnamefont {S.}~\bibnamefont {Hofferberth}},\ }\href@noop {} {\bibfield
  {journal} {\bibinfo  {journal} {Journal of Physics B: Atomic, Molecular and
  Optical Physics}\ }\textbf {\bibinfo {volume} {50}},\ \bibinfo {pages}
  {133001} (\bibinfo {year} {2017})}\BibitemShut {NoStop}%
\bibitem [{\citenamefont {Balewski}\ \emph {et~al.}(2014)\citenamefont
  {Balewski}, \citenamefont {Krupp}, \citenamefont {Gaj}, \citenamefont
  {Hofferberth}, \citenamefont {L\"ow},\ and\ \citenamefont
  {Pfau}}]{Balewski14-2}%
  \BibitemOpen
  \bibfield  {author} {\bibinfo {author} {\bibfnamefont {J.~B.}\ \bibnamefont
  {Balewski}}, \bibinfo {author} {\bibfnamefont {A.~T.}\ \bibnamefont {Krupp}},
  \bibinfo {author} {\bibfnamefont {A.}~\bibnamefont {Gaj}}, \bibinfo {author}
  {\bibfnamefont {S.}~\bibnamefont {Hofferberth}}, \bibinfo {author}
  {\bibfnamefont {R.}~\bibnamefont {L\"ow}}, \ and\ \bibinfo {author}
  {\bibfnamefont {T.}~\bibnamefont {Pfau}},\ }\href@noop {} {\bibfield
  {journal} {\bibinfo  {journal} {New Journal of Physics}\ }\textbf {\bibinfo
  {volume} {16}},\ \bibinfo {pages} {063012} (\bibinfo {year}
  {2014})}\BibitemShut {NoStop}%
\bibitem [{\citenamefont {van Bijnen}\ \emph {et~al.}(2011)\citenamefont {van
  Bijnen}, \citenamefont {Smit}, \citenamefont {van Leeuwen}, \citenamefont
  {Vredenbregt},\ and\ \citenamefont {Kokkelmans}}]{vanBijnen11}%
  \BibitemOpen
  \bibfield  {author} {\bibinfo {author} {\bibfnamefont {R.~M.~W.}\
  \bibnamefont {van Bijnen}}, \bibinfo {author} {\bibfnamefont
  {S.}~\bibnamefont {Smit}}, \bibinfo {author} {\bibfnamefont {K.~A.~H.}\
  \bibnamefont {van Leeuwen}}, \bibinfo {author} {\bibfnamefont {E.~J.~D.}\
  \bibnamefont {Vredenbregt}}, \ and\ \bibinfo {author} {\bibfnamefont {S.~J.
  J. M.~F.}\ \bibnamefont {Kokkelmans}},\ }\href@noop {} {\bibfield  {journal}
  {\bibinfo  {journal} {Journal of Physics B: Atomic, Molecular and Optical
  Physics}\ }\textbf {\bibinfo {volume} {44}},\ \bibinfo {pages} {184008}
  (\bibinfo {year} {2011})}\BibitemShut {NoStop}%
\bibitem [{\citenamefont {Robicheaux}\ and\ \citenamefont
  {Hern\'andez}(2005)}]{Robicheaux05}%
  \BibitemOpen
  \bibfield  {author} {\bibinfo {author} {\bibfnamefont {F.}~\bibnamefont
  {Robicheaux}}\ and\ \bibinfo {author} {\bibfnamefont {J.~V.}\ \bibnamefont
  {Hern\'andez}},\ }\href@noop {} {\bibfield  {journal} {\bibinfo  {journal}
  {Physical Review A}\ }\textbf {\bibinfo {volume} {72}},\ \bibinfo {pages}
  {063403} (\bibinfo {year} {2005})}\BibitemShut {NoStop}%
\bibitem [{\citenamefont {de~Hond}\ \emph {et~al.}(2018)\citenamefont
  {de~Hond}, \citenamefont {Cisternas}, \citenamefont {Spreeuw}, \citenamefont
  {van Linden van~den Heuvell},\ and\ \citenamefont {van
  Druten}}]{forthcomingpaper}%
  \BibitemOpen
  \bibfield  {author} {\bibinfo {author} {\bibfnamefont {J.}~\bibnamefont
  {de~Hond}}, \bibinfo {author} {\bibfnamefont {N.}~\bibnamefont {Cisternas}},
  \bibinfo {author} {\bibfnamefont {R.~J.~C.}\ \bibnamefont {Spreeuw}},
  \bibinfo {author} {\bibfnamefont {H.~B.}\ \bibnamefont {van Linden van~den
  Heuvell}}, \ and\ \bibinfo {author} {\bibfnamefont {N.~J.}\ \bibnamefont {van
  Druten}},\ }\href@noop {} {\bibfield  {journal} {\bibinfo  {journal} {in
  preparation}\ } (\bibinfo {year} {2018})}\BibitemShut {NoStop}%
\end{thebibliography}%

\end{document}